\documentclass[
 preprint,
 superscriptaddress,
 amsmath,amssymb,
 aps,
 prx,
 nofootinbib,
 showkeys
]{revtex4-1}

\clearpage{}\usepackage[uline]{hhtensor}
\usepackage{mathrsfs}
\DeclareMathOperator{\sign}{sign}

\newcommand\equaldueto[1]{\stackrel{#1}{=}}
\newcommand\alphaemp{\text{Re}\{\alpha^\omega_{me}\}}
\newcommand\alphatensor{\begin{bmatrix} \matr{\alpha}^\omega_\text{ee} & \matr{\alpha}^\omega_\text{em} \\
\matr{\alpha}^\omega_\text{me} & \matr{\alpha}^\omega_\text{mm} \end{bmatrix}}
\newcommand\Tnmp{\text{Re}\{T^\omega_{MN}\}}

\newcommand\Eq[1]{Eq.~(\ref{#1})}

\newcommand\Fourier[1]{\mathscr{F}\{#1\}}
\newcommand\FourierinvthreeD[1]{\mathscr{F}^{-1}_{3D}{\left\{#1\right\}}}

\newcommand\rr{\mathbf{r}}
\newcommand\rrbar{\mathbf{\bar{r}}}
\newcommand\zhat{\mathbf{\hat{z}}}
\newcommand\xhat{\mathbf{\hat{x}}}
\newcommand\yhat{\mathbf{\hat{y}}}

\newcommand\pp{\mathbf{k}}

\newcommand\phat{\mathbf{\hat{\pp}}}
\newcommand\ehatplusminus{\mathbf{\hat{e}}_{\pm}(\phat)}

\newcommand\Vtall{\mathcal{V}}
\newcommand\Vpplus{\mathbf{V}_+(\pp)}
\newcommand\Vrtplus{\mathbf{V}_+(\rr,t)}
\newcommand\Vrtplusminus{\mathbf{V}_\pm(\rr,t)}
\newcommand\Vrwplus{\mathbf{V}^\omega_+(\rr)}
\newcommand\Vpminus{\mathbf{V}_-(\pp)}
\newcommand\Vpplusminus{\mathbf{V}_\pm(\pp)}
\newcommand\Vrtminus{\mathbf{V}_-(\rr,t)}
\newcommand\Vrwminus{\mathbf{V}^\omega_-(\rr)}
\newcommand\Vrwplusminus{\mathbf{V}^\omega_\pm(\rr)}

\newcommand\omegabar{\bar{\omega}}

\newcommand\Fket{|\Ftall\rangle}
\newcommand\Fbarket{|\Fbartall\rangle}
\newcommand\Ftall{\mathcal{G}}
\newcommand\Fbartall{\mathcal{\bar{G}}}

\newcommand\Gket{|\Gtall\rangle}
\newcommand\Gbarket{|\Gbartall\rangle}
\newcommand\Gtall{\mathcal{F}}
\newcommand\Gbartall{\mathcal{\bar{F}}}
\newcommand\Gpplus{\mathbf{F}_+(\pp)}
\newcommand\Grtplus{\mathbf{F}_+(\rr,t)}
\newcommand\Grbartplus{\mathbf{F}_+(\rrbar,t)}
\newcommand\Grtplusminus{\mathbf{F}_{\pm}(\rr,t)}
\newcommand\Grwplus{\mathbf{F}^\omega_+(\rr)}
\newcommand\Grwbarplus{\mathbf{F}^{\omegabar}_+(\rr)}
\newcommand\Gpminus{\mathbf{F}_-(\pp)}
\newcommand\Grtminus{\mathbf{F}_-(\rr,t)}
\newcommand\Grbartminus{\mathbf{F}_-(\rrbar,t)}
\newcommand\Grwminus{\mathbf{F}^\omega_-(\rr)}
\newcommand\Grwplusminus{\mathbf{F}^\omega_\pm(\rr)}
\newcommand\Grwbarminus{\mathbf{F}^{\omegabar}_-(\rr)}
\newcommand\Gpplusminus{\mathbf{F}_\pm(\pp)}

\newcommand\Ert{\mathbf{E}(\rr,t)}
\newcommand\Drt{\mathbf{D}(\rr,t)}
\newcommand\Brt{\mathbf{B}(\rr,t)}
\newcommand\Hrt{\mathbf{H}(\rr,t)}
\newcommand\Art{\mathbf{A}(\rr,t)}
\newcommand\Crt{\mathbf{C}(\rr,t)}

\newcommand\Erealrt{\mathcal{E}(\rr,t)}
\newcommand\Etransrealrt{\mathcal{E}_{\perp}(\rr,t)}
\newcommand\Drealrt{\mathcal{D}(\rr,t)}
\newcommand\Brealrt{\mathcal{B}(\rr,t)}
\newcommand\Arealrt{\mathcal{A}(\rr,t)}
\newcommand\Crealrt{\mathcal{C}(\rr,t)}

\newcommand\Btildert{\mathcal{{B}}(\rr,t)}
\newcommand\Dtildert{\mathcal{{D}}(\rr,t)}

\newcommand\Brtreal{\mathbf{B}_{\text{re}}(\rr,t)}
\newcommand\Brtimag{\mathbf{B}_{\text{im}}(\rr,t)}
\newcommand\Artreal{\mathbf{A}_{\text{re}}(\rr,t)}
\newcommand\Artimag{\mathbf{A}_{\text{im}}(\rr,t)}
\newcommand\Crtreal{\mathbf{C}_{\text{re}}(\rr,t)}
\newcommand\Crtimag{\mathbf{C}_{\text{im}}(\rr,t)}
\newcommand\Drtreal{\mathbf{D}_{\text{re}}(\rr,t)}
\newcommand\Drtimag{\mathbf{D}_{\text{im}}(\rr,t)}

\newcommand\Xrw{\mathbf{X}^\omega(\rr)}
\newcommand\Xrt{\mathbf{X}(\rr,t)}
\newcommand\Xtildert{\mathcal{{X}}(\rr,t)}

\newcommand\Xrtreal{\mathbf{X}_{\text{re}}(\rr,t)}
\newcommand\Xrtimag{\mathbf{X}_{\text{im}}(\rr,t)}

\newcommand\intdpconf{\int_{\mathbb{R}^3-\{\mathbf{0}\}} \frac{d \pp}{c_0|\pp|}\text{ }}
\newcommand\intdp{\int_{\mathbb{R}^3-\{\mathbf{0}\}} d \pp\text{ }}
\newcommand\intdpnorm{\int_{\mathbb{R}^3-\{\mathbf{0}\}} \frac{d \pp}{\sqrt{(2\pi)^3}}\text{ }}
\newcommand\intdomega{\int_{>0}^\infty \frac{d \omega}{\sqrt{2\pi}}\text{ }}
\newcommand\intdomegatwo{\int_{-\infty}^\infty \frac{d \omega}{\sqrt{2\pi}}\text{ }}
\newcommand\intdomegatwonopi{\int_{-\infty}^\infty d \omega\text{ }}
\newcommand\intdomegabar{\int_{>0}^\infty \frac{d \omegabar}{\sqrt{2\pi}}\text{ }}

\newcommand\intdomeganopi{\int_{>0}^\infty d \omega\text{ }}

\newcommand\intdr{\int_{\mathbb{R}^3} {d \rr}\text{ }}
\newcommand\intdrbar{\int_{\mathbb{R}^3} {d \bar{\rr}}\text{ }}
\newcommand\intdrD{\int_{\mathbb{D}} {d \rr}\text{ }}
\newcommand\AVLambda{\langle \Gtall|\Lambda|\Gtall \rangle}
\newcommand\AVGamma{\langle \Gtall|\Gamma|\Gtall \rangle}
\clearpage{}
\begin{document}
\clearpage{}\title{A Conformally Invariant Derivation of Average Electromagnetic Helicity}
\author{Ivan Fernandez-Corbaton}
\affiliation{Institute of Nanotechnology, Karlsruhe Institute of Technology, 76021 Karlsruhe, Germany}
\email{ivan.fernandez-corbaton@kit.edu}
\begin{abstract}
The average helicity of a given electromagnetic field measures the difference between the number of left- and right-handed photons contained in the field. Here, the average helicity is derived using the conformally invariant inner product for Maxwell fields. Several equivalent integral expressions in momentum space, in $(\rr,t)$ space, and in the time-harmonic $(\rr,\omega)$ space are obtained, featuring Riemann–Silberstein-like fields and potentials. The time-harmonic expressions can be directly evaluated using the outputs of common numerical solvers of Maxwell equations. The results are shown to be equivalent to the well-known volume integral for the average helicity, featuring the electric and magnetic fields and potentials.
 \end{abstract}
\maketitle
\clearpage{}
\clearpage{}
The helicity of electromagnetic fields has received research attention since at least the 1960s \cite{Calkin1965,Zwanziger1968,Deser1976,Birula1981,Birula1996,Afanasiev1996,Trueba1996,Drummond1999}. Recently, the topic has picked up a considerable pace, partly because of the relevance of helicity in chiral light–matter interactions \cite{Coles2012,FerCor2012p,Cameron2012,Bliokh2013,Cameron2013,Nieto2015,Gutsche2016,FerCor2016,Elbistan2017,Andrews2018,Vazquez2018,Hanifeh2018b,Crimin2019,FernandezGuasti2019,Poulikakos2019,Bernabeu2019}. In this context, one of the most basic quantities is the average value of electromagnetic helicity for a given free electromagnetic field, which can be interpreted as the pseudo-scalar that measures the difference between the number of left- and right-handed polarized photons contained in the field \cite{Calkin1965,Zwanziger1968}. The most common expression for average helicity is \cite{Calkin1965,Zwanziger1968,Deser1976,Afanasiev1996,Trueba1996,Drummond1999,Cameron2012,Bliokh2013,Cameron2013,Nieto2015,Gutsche2016,Elbistan2017,Crimin2019,FernandezGuasti2019,Poulikakos2019,Bernabeu2019}:
\begin{equation*}
	\frac{1}{2}\intdr \Brealrt\cdot\Arealrt-\Erealrt\cdot\Crealrt,
\end{equation*}
where $\Erealrt\left[\Crealrt\right]$ and $\Brealrt\left[\Arealrt\right]$ are the real-valued electric and magnetic fields[potentials], respectively. The use of two potentials is a common strategy in this context \cite{Zwanziger1968,Drummond1999,Cameron2012,Cameron2013,Bliokh2013,Nienhuis2016,Elbistan2017}. In particular, it allows the obtaining of an integrand which is local in $\rr$. The above equation has been arrived at in several ways. This article contains a different one.

In this article, several different integral expressions for the average helicity of a given electromagnetic field are obtained starting from the inner product that ensures invariance of the result under the largest symmetry group of Maxwell equations: The conformal group. We show that the result from the derived expressions coincides with that of the above integral. In our approach, all the fields are complex because only positive frequencies are included. The advantages of this choice regarding the treatment of helicity with Riemann-Silberstein-like fields and their corresponding potentials are discussed. Integral expressions in momentum space, $(\rr,t)$ space, and $(\rr,\omega)$ space are obtained. The numerical evaluation of the time-harmonic $(\rr,\omega)$ expressions, whose integrands are local in $\rr$, can be conveniently performed using common numerical Maxwell solvers. The formalism is used to obtain expressions for the computation of molecular circular dichroism in Appendix~\ref{app:cd}. 

We start by writing down the conformally invariant form of the average helicity and showing that the result coincides with the common definition. To such end, we consider $\mathbb{M}$, the vector space of finite-energy solutions to Maxwell's equations in free space. We denote vectors in such space by kets such as $\Gket$, which represent particular electromagnetic field solutions. We are interested in the average helicity of the field. The key ingredient for the definition of average properties is an inner product between two vectors $\Gket$ and $|\Ftall\rangle$, denoted $\langle \Ftall\Gket$, which endows $\mathbb{M}$ with the structure of a Hilbert space. Properties such as energy, linear momentum, helicity, etc. ..., are represented by Hermitian operators which map elements of $\mathbb{M}$ back onto itself. Then, for a given field $\Gket$, the average value of a given property represented by a Hermitian operator $\Gamma$ is the quadratic form
\begin{equation}
	\label{eq:inner}
\langle \Gtall|\Gamma|\Gtall \rangle, 
\end{equation}
that is, the projection of the vector $\Gket$ onto the vector $\Gamma\Gket$. Since $\Gamma$ is Hermitian, $\langle \Gtall|\Gamma|\Gtall \rangle$ must be a real number.

The crucial question of {\em which inner product} to choose was settled by Gross by requiring the inner product to be invariant under the conformal group \cite{Gross1964}. That is: Given any two solutions $\Gket$ and $\Fket$, and their corresponding transformed versions under any transformation in the conformal group, $\Gbarket$ and $\Fbarket$, the inner product must be so that $\langle \Gtall|\Ftall\rangle=\langle \Gbartall|\Fbartall\rangle$. Gross showed in Ref. \cite{Gross1964} that this requirement essentially determines the exact expression of the inner product, which we will use later. The conformal group includes space–time translations, spatial rotations, and Lorentz boosts, which~together form the Poincar\'e group, plus space–time scalings, and special conformal transformations \cite{Gross1964}. The conformal group is the largest symmetry group of Maxwell equations in free space. Conformally invariant results have hence the maximum possible validity in electromagnetism. 

An important distinction is in order at this point. The use of the conformally invariant inner product ensures the maximal validity for average quantities as defined by \Eq{eq:inner}: The projection of the vector $|\mathcal{F}\rangle$ onto the vector $\Gamma|\mathcal{F}\rangle$ is equal to the projection of $\Gbarket=T|\mathcal{F}\rangle$ onto $|\overline{\Gamma\mathcal{F}}\rangle=T\Gamma|\mathcal{F}\rangle$, i.e., $\langle \mathcal{F}|T^\dagger T\Gamma|\mathcal{F}\rangle=\langle \mathcal{F}|\Gamma|\mathcal{F}\rangle$, for any transformation $T$ in the conformal group, where ${T}^\dagger$ is the Hermitian adjoint of $T$. Satisfying this demand amounts to showing that an inner product exists with respect to which the conformal group acts unitarily ($TT^\dagger=T^\dagger T=I$ for all $T$, where $I$ is the identity) on the vector space of solutions of Maxwell equations \cite{Gross1964}. Loosely speaking, this means that the value of the averages in \Eq{eq:inner} will not change regardless of ``the conformal point of view'' or ``conformal change of coordinate system''. This will hold for average helicity, and also for average momentum, average angular momentum, etc. \ldots. A {\em different} matter is whether the average quantity in a conformally transformed field is the same as the average quantity in the initial field, for all conformal transformations. In this case, we are asking whether $\Gket$ and $T\Gket$ have the same average value of a given property $\Gamma$, i.e., whether 
\begin{equation}
	\label{eq:mess}
	\langle \mathcal{F}|T^\dagger \Gamma T|\mathcal{F}\rangle=\langle \mathcal{F}|\Gamma|\mathcal{F}\rangle \text{ for all $T$},
\end{equation}
	which is often not the case, such as for example when a Lorentz boost simultaneously changes the energy and momentum of a given field. Incidentally, it will be clear later that \Eq{eq:mess} is actually met in the case of average helicity. 

Writing down an explicit expression for \Eq{eq:inner} requires us to choose an explicit representation for the vectors in $\mathbb{M}$ and the operators acting on them. We choose the following representation for the vectors in $\mathbb{M}$:
\begin{equation}
	\label{eq:rep}
	\Gket \equiv \Gtall(\pp)=\begin{bmatrix}\Gpplus\\\Gpminus\end{bmatrix},
\end{equation}
where the $\Gpplusminus$ define the plane-wave components of a version of the Riemann-Silberstein vectors \cite{Birula1996} 
\begin{equation}
	\label{eq:gp}
	\begin{split}
		\frac{\Drt}{\sqrt{2\epsilon_0}} \pm i\frac{\Brt}{\sqrt{2\mu_0}}&= \sqrt{\frac{\epsilon_0}{2}}\left[\Ert\pm iZ_0\Hrt\right]=\Grtplusminus\\&=\intdpnorm \Gpplusminus\exp(i\pp\cdot\rr-i\omega t)\text{, with $c_0\sqrt{\pp\cdot\pp}=c_0k=\omega>0$,}
	\end{split}
\end{equation}
where $\epsilon_0$, $\mu_0$, $c_0$, and $Z_0=\sqrt{\mu_0/\epsilon_0}$ are the vacuum's permittivity, permeability, speed of light, and impedance, respectively, $\pp$ is the wavevector, and $\omega=c_0k=c_0\sqrt{\pp\cdot\pp}$ is the angular frequency. The $\Gpplusminus$ can be further decomposed as $\Gpplusminus=\ehatplusminus f_\pm(\pp)$, where $f_\pm(\pp)$ are complex-valued scalar functions and $\ehatplusminus$ are the $\phat$-dependent polarization vectors\footnote{The $\ehatplusminus$ can be obtained by the rotation of $(\pm\xhat-i\yhat)/\sqrt{2}$, the two vectors corresponding to $\phat=\zhat$: $\sqrt{2}\ehatplusminus=R_z(\phi)R_y(\theta)\left(\mp \xhat -i\yhat\right)$, where $\theta=\arccos\left(k_z/k\right)$ and $\phi=\arctan\left(k_y,k_x\right)$.} for each handedness(helicity). We note that $\phat\cdot\ehatplusminus=0$, which makes the $\Gpplusminus[\Grtplusminus)]$ transverse functions, namely, $\phat\cdot\Gpplusminus=\nabla\cdot\Grtplusminus=0$. The origin $\pp=\mathbf{0}$ is removed from the integral in \Eq{eq:gp} because we are considering electrodynamics and excluding electro- and magneto-statics, whereby $k=\omega/c_0=0$ needs to be excluded.

It important to note that {\em only positive frequencies are included} in \Eq{eq:gp}. This amounts to considering positive energies only, which is possible in electromagnetism since the photon is its own anti-particle. Only one sign of the energy(frequency) is needed because the same information is contained on both sides of the spectrum  \cite[\S 3.1]{Birula1996}\cite{Birula1981}. When only positive frequencies are included, $\Drt$, $\Brt$, $\Ert$ and $\Hrt$ in \Eq{eq:gp} are complex-valued fields. With $\mathbf{X}$ standing for $\mathbf{D}$, $\mathbf{B}$, $\mathbf{E}$ or $\mathbf{H}$:
\begin{equation}
	\label{eq:tomega}
	\Xrt=\intdomega \Xrw\exp(-i\omega t).
\end{equation}
We define the complex-valued fields so that the typical real-valued versions are obtained as
\begin{equation}
	\label{eq:real}
	\Xtildert=\intdomega \Xrw\exp(-i\omega t)+{\Xrw}^*\exp(i\omega t).
\end{equation}
The restriction to positive frequencies is particularly consequential for the treatment of helicity, the generalized polarization handednesses of the field. One of the advantages of the Riemann-Silberstein vectors is their ability to encode the helicity content of the field. They are the eigenstates of the helicity operator and potentially allow for the splitting of the two polarization handedness in any field, including near and evanescent fields. However, when they are defined by means of real-valued fields, as in $\frac{\Dtildert}{\sqrt{2\epsilon_0}} \pm i\frac{\Btildert}{\sqrt{2\mu_0}}$, their use for splitting the two helicities is not as simple as it becomes when complex-valued fields are used. With real-valued fields we have that the two $\pm$ fields determine each other through complex conjugation $\left[{\frac{\Dtildert}{\sqrt{2\epsilon_0}}+i\frac{\Btildert}{\sqrt{2\mu_0}}}\right]^*={\frac{\Dtildert}{\sqrt{2\epsilon_0}}-i\frac{\Btildert}{\sqrt{2\mu_0}}}$, which is at odds with the a priori physical independence of the two helicity components of the electromagnetic field. For example, the complex conjugation connection means that the two $\pm$ squared norms $\left|{\frac{\Dtildert}{\sqrt{2\epsilon_0}}\pm i\frac{\Btildert}{\sqrt{2\mu_0}}}\right|^2$, which could intuitively be thought of as the $(\rr,t)$-local helicity intensities, become equal at all space-time points $\left|{\frac{\Dtildert}{\sqrt{2\epsilon_0}}+i\frac{\Btildert}{\sqrt{2\mu_0}}}\right|^2=\left|{\frac{\Dtildert}{\sqrt{2\epsilon_0}}-i\frac{\Btildert}{\sqrt{2\mu_0}}}\right|^2$. This contradicts, for example, the fact that there can be electromagnetic fields containing only one of the two helicities, e.g., any linear combination of plane-waves with the same polarization handedness. The restriction to positive frequencies overcomes these limitations: In \Eq{eq:gp}, $\Grtplus$ contains no information about $\Grtminus$, in particular ${\Grtplus}^*\neq\Grtminus$. 

The choice of the representation in \Eq{eq:rep}, where the two helicity components are distinguished, as opposed to other more common possibilities where the electric and magnetic fields are distinguished, can also be motivated by the transformation properties of the two different options with respect to the conformal group. Namely helicity is invariant under the conformal group \cite{Mack1969,Mack1977}. That is, all the generators of the conformal group commute with the helicity operator. This was established by Mack and Todorov in Ref.~\onlinecite{Mack1969} when they showed that a Casimir operator of the conformal group is linearly related to the helicity operator. The invariance can also be inferred from the facts that (i) helicity is ultimately proportional to the cosine of an angle\footnote{The cosine of the angle between the vector of spin-1 matrices $\mathbf{S}$ and the linear momentum operator $\mathbf{P}$ is defined as $\cos\left[\angle\left(\mathbf{S},\mathbf{P}\right)\right]=\frac{\mathbf{S}\cdot\mathbf{P}}{|\mathbf{P}||\mathbf{S}|}$. Then, using the definition of the helicity operator $\Lambda$ in \Eq{eq:heldef} we can write $\Lambda=\cos\left[\angle\left(\mathbf{S},\mathbf{P}\right)\right]|\mathbf{S}|$. But the action of $|\mathbf{S}|$ on members of $\mathbb{M}$ is trivial since $|\mathbf{S}|^2=\left(S_1^2+S_2^2+S_3^2\right)=2I$ where $I$ is the identity. This can be seen in \cite[Eq.~(5.54)]{Rose1957}, and is readily verified by direct calculation using the spin-1 matrices.}, and that, (ii) the preservation of angles is guaranteed by conformal transformations. In the representation of \Eq{eq:rep} this invariance means that no matter which conformal transformation is applied to $\Gket$, the $\Gpplus$ upper components of $\Gtall(\pp)$ will never end up on the lower part, and vice-versa. This reduces the algebraic complexity of some expressions and manipulations. In sharp contrast to this, what is meant by electric and magnetic fields is not conformally invariant. Actually, the meanings of ``electric'' and ``magnetic'' are not even relativistically invariant since electric and magnetic fields are intermixed by Lorentz boosts \cite[Eq. (11.149)]{Jackson1998}. An important physical fact about helicity can be deduced from its conformal invariance. Since the helicity operator ($\Lambda$) commutes with any transformation $T$ in the conformal group ($T\Lambda=\Lambda T$), and $T$ is unitary with respect to the chosen inner product ($TT^\dagger =T^\dagger T=I$), we can readily see that \Eq{eq:mess} is met\footnote{$\langle \mathcal{F}|T^\dagger \Lambda T|\mathcal{F}\rangle\equaldueto{\Lambda T=T\Lambda}\langle \mathcal{F}|T^\dagger T\Lambda |\mathcal{F}\rangle\equaldueto{T^\dagger T=I} \langle\mathcal{F}|\Lambda|\mathcal{F}\rangle$.}: The average helicity of a conformally transformed field is the same as the average helicity of the initial field.

Let us go on to computing the average helicity of a given field as a conformally invariant inner product. We will explicitly keep the constants $\epsilon_0$, $\mu_0$, $c_0$, and $Z_0$ in the expressions, and use the four fields $\mathbf{D}$, $\mathbf{B}$, $\mathbf{E}$ and $\mathbf{H}$. These choices \cite{Crimin2019,Poulikakos2019} facilitate the re-use of the formulas when a description such as the one in \Eq{eq:gp} is possible in a non-vacuum background, such as for example in an infinite isotropic and homogeneous linear medium. 

Following Gross \cite{Gross1964}, and Bialynicki-Birula \cite[\S 9]{Birula1975}\cite[\S 5]{Birula1996}, the definitions in Eqs.~(\ref{eq:inner},\ref{eq:rep}) allow us to write the average value of any property $\Gamma$ as\footnote{This is seen by comparing \Eq{eq:gp} with \cite[Eqs. (4.11)-(4.12)]{Birula1996}, and \Eq{eq:first} particularized to the energy operator $\Gamma\rightarrow H=\begin{bmatrix}\omega I_{3\times 3}&0_{3\times 3}\\0_{3\times 3}&\omega I_{3\times 3}\end{bmatrix}$ with \cite[Eq. (4.13)]{Birula1996}, and setting $\hbar=1$.}
\begin{equation}
	\label{eq:first}
\AVGamma= \intdpconf {\Gtall(\pp)}^\dagger\Gamma\Gtall(\pp),
\end{equation}
where ${}^\dagger$ means transpose conjugate. Equation~(\ref{eq:first}) is an explicit expression of the conformally invariant inner product between $\Gket$ and $\Gamma\Gket$\footnote{When \Eq{eq:first} is brought to the $(\rr,t)$ domain, the $\pp$-local expression results in the double integral $\int_{\mathbb{R}^3} d\rr\int_{\mathbb{R}^3} d\bar{\mathbf{r}}$ of a manifestly non-$\rr$-local integrand including a term like $1/|\rr-\mathbf{\bar{r}}|^2$ (see \cite[Eq.~(6)]{Gross1964} and \cite[Eq.~(5.7)]{Birula1996}).}.

We are now ready to focus our attention on the average value of helicity. The helicity operator $\Lambda$ is defined as the projection of the angular momentum operator vector $\mathbf{J}$ onto the direction of the linear momentum operator vector $\mathbf{P}$:
\begin{equation}
\label{eq:heldef}
\Lambda=\frac{\mathbf{J}\cdot\mathbf{P}}{|\mathbf{P}|}=\frac{\mathbf{S}\cdot\mathbf{P}}{|\mathbf{P}|},
\end{equation}
where for electromagnetism, $\mathbf{S}$ is the vector of spin-1 matrices\footnote{The second equality can be seen to follow, for example, from considering the coordinate representation of the angular momentum and linear momentum operator vectors, \cite[Eqs.~(5.24,5.25)]{Birula1996}: $\mathbf{J}\equiv -i\rr\times\nabla + \mathbf{S}$, $\mathbf{P}\equiv -i\nabla$. Their inner product then reads $\mathbf{J}\cdot\mathbf{P}\equiv -(\rr\times\nabla)\cdot\nabla-i \mathbf{S}\cdot\nabla$. The first term vanishes since it is the divergence of a curl.}. 

We start by particularizing \Eq{eq:first} to the helicity operator $\Lambda$.
\begin{equation}
	\label{eq:Hel}
\AVLambda	 = \intdpconf {\Gtall(\pp)}^\dagger\Lambda\Gtall(\pp)=\intdpconf \begin{bmatrix}\Gpplus\\\Gpminus\end{bmatrix}^\dagger\begin{bmatrix}i\phat\times&0\\0&i\phat\times\end{bmatrix}\begin{bmatrix}\Gpplus\\\Gpminus\end{bmatrix},
\end{equation}
where the last expression contains the explicit form of the helicity operator in our choice of representation\footnote{This follows from the definition of helicity in \Eq{eq:heldef}: $\mathbf{S}\cdot\mathbf{P}/|\mathbf{P}|\equiv i\phat\times$, where the equivalence follows from applying \cite[Eq. (2.2)]{Birula1996} in momentum space where $\mathbf{P}\rightarrow \pp\implies\mathbf{P}/|\mathbf{P}|\rightarrow \phat$. }. We now use the fact that the $\Gpplusminus$ are eigenstates of helicity, namely $i\phat\times \Gpplusminus =\pm \Gpplusminus$, to write
\begin{equation}
	\label{eq:AVGtall}
	\AVLambda= \intdpconf \Gpplus^\dagger\Gpplus-\Gpminus^\dagger\Gpminus=\intdpconf |\Gpplus|^2-|\Gpminus|^2.
\end{equation}
We will now show that \Eq{eq:AVGtall} is equivalent to the most common integral expression of the helicity average. To such end, and taking advantage of the fact that $k\neq 0$, we define the helicity potentials 
\begin{equation}
	\label{eq:V}
	\Vtall(\pp)=\frac{1}{ikc_0}\Gtall(\pp)=\frac{1}{ikc_0}\begin{bmatrix}\Gpplus\\\Gpminus\end{bmatrix}=\begin{bmatrix}\Vpplus\\\Vpminus\end{bmatrix},
\end{equation}
which in the $(\rr,t)$ representation, and recalling that $-i\omega\rightarrow\partial_t$, are seen to be related to $\Grtplusminus$ as
\begin{equation}
	\label{eq:GA}
	-\partial_t \Vrtplusminus=\Grtplusminus=\sqrt{\frac{\epsilon_0}{2}}\left[\Ert\pm iZ_0\Hrt\right],
\end{equation}
from where we can use \cite[Eq.~(2)]{Crimin2019}, namely $-\partial_t\Crt=\Hrt\text{ and }-\partial_t\Art=\Ert$, to recognize that these helicity potentials are linear combinations of complex versions of the transverse real-valued ``magnetic'' $\Arealrt$ and ``electric'' $\Crealrt$ potentials \cite{Zwanziger1968,Drummond1999,Cameron2012,Cameron2013,Bliokh2013,Nienhuis2016,Elbistan2017}.
\begin{equation}
	\label{eq:AC}
	\Vrtplusminus=\sqrt{\frac{\epsilon_0}{2}}\left[\Art\pm iZ_0\Crt\right].
\end{equation}
Appendix~\ref{app:second} contains some background information about the electric potential. Linear combinations very similar to \Eq{eq:AC} have been recently introduced by Elbistan {\em et. al} in Ref.~\onlinecite{Elbistan2017}, albeit using real-valued vector functions instead of our complex $\Art$ and $\Crt$. As previously discussed, this difference is relevant for treating helicity. When real-valued fields are used in the right hand side of \Eq{eq:AC}, it follows that ${\Vrtplus}^*=\Vrtminus$, which ultimately leads to a zero value of the average helicity as reported in \cite{Elbistan2017}.

It is also worth pointing out that the $\Vpplusminus$ functions are transverse, i.e., $\phat\cdot\Vpplusminus=0$, which follows from \Eq{eq:V} and the previously mentioned property $\phat\cdot\Gpplusminus$. The helicity potentials only contain the transverse degrees of freedom, the same as the free electromagnetic field, which ensures that the results obtained using $\Vpplusminus$ are gauge independent. 

We proceed by using \Eq{eq:V} and the central expression in \Eq{eq:AVGtall} to obtain 
\begin{equation}
	\label{eq:helv}
		\AVLambda=i \intdp \Gpplus^\dagger \Vpplus - \Gpminus^\dagger \Vpminus.
\end{equation}
Equation~(\ref{eq:helv}) can now be brought to the $(\rr,t)$ domain as follows. First, we apply the substitutions $\Gpplusminus\rightarrow\Gpplusminus\exp(-ic_0k)$, and $\Vpplusminus\rightarrow\Vpplusminus\exp(-ic_0k)$
\begin{equation}
	\label{eq:helvnew}
	\begin{split}
		&\AVLambda=\\
		&i \intdp \left[\Gpplus\exp(-ic_0k)\right]^\dagger \left[\Vpplus\exp(-ic_0k)\right] - \left[\Gpminus\exp(-ic_0k)\right]^\dagger \left[\Vpminus\exp(-ic_0k)\right].
	\end{split}
\end{equation}
These changes do not affect the result, but allow us to see from \Eq{eq:gp} that the $\Gpplusminus\exp(-ic_0k)$ are the three-dimensional Fourier transforms ($\rr\rightarrow\pp$) of $\Grtplusminus$. The same relation holds between $\Vpplusminus\exp(-ic_0k)$ and $\Vrtplusminus$. We can now apply applying Parseval's theorem, i.e., the unitarity of the inverse Fourier transform $\pp\rightarrow\rr$, to each of the two terms in the subtraction in \Eq{eq:helvnew}:
\begin{equation}
	\label{eq:helgv}
	\AVLambda=i\intdr \Grtplus^\dagger\Vrtplus -\Grtminus^\dagger\Vrtminus,
\end{equation}
where the integrand is local in $\rr$. We show in Appendix~\ref{app:second} that when \Eq{eq:AVGtall} is brought to the $(\rr,t)$ domain instead, the $1/|\pp|$ term results in the double integral $\int_{\mathbb{R}^3} d\rr\int_{\mathbb{R}^3} d\bar{\mathbf{r}}$ of a manifestly non-$\rr$-local integrand including a term such as $1/|\rr-\mathbf{\bar{r}}|^2$. The inconvenient $1/|\pp|$ term is absorbed in the definition of the potentials in \Eq{eq:V}.

To further approach the most common expression of the average helicity, we now substitute 
\begin{equation}
	\mathbf{F}_{\pm}(\rr,t) =\frac{\Drt}{\sqrt{2\epsilon_0}}\pm i\frac{\Brt}{\sqrt{2\mu_0}},\ 
	\mathbf{V}_{\pm}(\rr,t)= \sqrt{\frac{\epsilon_0}{2}}\left[\Art\pm iZ_0\Crt\right], 
\end{equation}
into \Eq{eq:helgv} and obtain
\begin{equation}
	\label{eq:nyaca}
	\begin{split}
		\AVLambda&=\frac{i}{2}\intdr\\
		 &\left[\Drt^\dagger \Art+iZ_0\Drt^\dagger\Crt-\frac{i}{Z_0}\Brt^\dagger\Art+\Brt^\dagger\Crt \right]\\
		 -&\left[\Drt^\dagger \Art-iZ_0\Drt^\dagger\Crt+\frac{i}{Z_0}\Brt^\dagger\Art+\Brt^\dagger\Crt \right]\\
		&=\intdr \frac{1}{Z_0}\Brt^\dagger\Art-Z_0\Drt^\dagger\Crt,
	\end{split}
\end{equation}
which is a complex version of the well-known integral for the average helicity featuring real-valued fields, as found e.g., in \cite[Eq.~(6)]{Crimin2019}. Appendix~\ref{app:complexreal} shows that the results of the complex and real versions coincide.

The $\pp$-domain expressions in Eqs.~(\ref{eq:AVGtall}) and (\ref{eq:helv}), and the $(\rr,t)$-domain expression in \Eq{eq:helgv} produce the correct result. We now obtain $(\rr,\omega)$-domain expressions. The time-harmonic decomposition is often used in both theoretical investigations and numerical computations. 

We start by noting that the result of the integral in Eq.~(\ref{eq:helgv}) is independent of time\footnote{Indeed, the simplifying arbitrary choice $t=0$ is made by Gross in \cite{Gross1964} for evaluating the inner product with integrals featuring $(\rr,t)$-dependent integrands.}. The time independence of $\AVLambda$, manifest in Eqs.~(\ref{eq:AVGtall}) and (\ref{eq:helv}), is ultimately due the fact that $\AVLambda$ must be invariant under time translations since such transformations are contained in the conformal group. That is, $\AVLambda$ cannot depend on time. This can be exploited to obtain expressions for $\AVLambda$ involving the time-harmonic decomposition of the fields. To such end, we go back to \Eq{eq:helgv}, and expand each term in the integrand into their frequency components
\begin{equation}
	\label{eq:jarl}
	\begin{split}
		\AVLambda&=i\intdr\\
		&\left[\intdomega \Grwplus\exp(-i\omega t)\right]^\dagger\left[\intdomega \Vrwplus\exp(-i\omega t)\right]\\
		-&\left[\intdomega \Grwminus\exp(-i\omega t)\right]^\dagger\left[\intdomega \Vrwminus\exp(-i\omega t)\right]\\
		&=i\intdr\intdomega\intdomegabar\left[\Grwbarplus^\dagger\Vrwplus-\Grwbarminus^\dagger\Vrwminus\right]\exp\left(-i(\omega-\omegabar)t\right).
	\end{split}
\end{equation}
Let us examine the last line of \Eq{eq:jarl}. Because $\AVLambda$ cannot depend on time, and since the two helicities are independent of each other, it follows that only the $\omega=\omegabar$ components can contribute to the end result. This allows us to obtain the following three equivalent expressions:
\begin{equation}
	\label{eq:lots}
	\begin{split}
		\left(2\pi\right)\AVLambda=&i\intdr\intdomeganopi\Grwplus^\dagger\Vrwplus-\Grwminus^\dagger\Vrwminus\\
					=&\intdr\intdomeganopi\frac{1}{\omega}\left(|\Grwplus|^2-|\Grwminus|^2\right)\\
					=&\intdr\intdomeganopi\omega\left(|\Vrwplus|^2-|\Vrwminus|^2\right),
	\end{split}
\end{equation}
where the equalities readily follow from $\Grwplusminus=i\omega\Vrwplusminus$, which follows from \Eq{eq:GA}.

Expressions that are local in $\rr$, such as Eqs.~(\ref{eq:helv}) and (\ref{eq:lots}), justify the consideration of the average helicity in a finite volume $\mathbb{D}$. This then allows use of the corresponding expressions in practical situations where numerical solvers calculate the fields in finite regions of space. The expressions in \Eq{eq:lots} are particularly adapted to the output of finite-element-method solvers such as COMSOL and JCM, which use the time-harmonic decomposition of $\rr$-dependent fields. 

Finally, regarding applications, the electromagnetic helicity is particularly relevant in chiral light-matter interactions. Among these, the interaction of the field with chiral molecules is one of the most researched cases, partly because the optical sensing of chiral molecules is important in chemistry and pharmaceutical applications. In Appendix~\ref{app:cd} we use the above formalism to derive expressions for computing the circular dichroism signal for two different settings of the light-molecule interaction: The $6\times 6$ dipolarizability tensor and the T-matrix.

In conclusion, several equivalent expressions for the average value of the electromagnetic helicity of a given field have been obtained from a starting point featuring maximal electromagnetic invariance, i.e., from an expression whose result is invariant under the conformal group. Some of the obtained expressions can be conveniently evaluated using the outputs of common Maxwell solvers.
\clearpage{}
\begin{acknowledgements}
	Partially funded by the Deutsche Forschungsgemeinschaft (DFG, German Research Foundation) --Project-ID 258734477-- SFB 1173. I would also like to acknowledge support by KIT through the Virtual Materials Design (VIRTMAT) project by the Helmholtz Association via the Helmholtz program Science and Technology of Nanosystems (STN). 
\end{acknowledgements}

\appendix
\clearpage{}\section{The use of two potentials for obtaining $\rr$-local integrands \label{app:second}}
The use of two potentials, one magnetic and one electric, has a long tradition in the studies of helicity and of the symmetry generated by the helicity operator: Electromagnetic duality \cite{Zwanziger1968,Drummond1999,Cameron2012,Cameron2013,Bliokh2013,Nienhuis2016,Elbistan2017}. Duality can be seen as the underlying reason for adding an electric potential next to the magnetic one. 

In the absence of sources, the (real-valued) electric potential $\Crealrt$ is typically defined by first fixing its transverse part 
\begin{equation}
	\label{eq:defC}
\Etransrealrt=-\nabla\times\Crealrt,
\end{equation}
where $\Etransrealrt$ is the transverse electric field, and then exploiting the fact that $\Crealrt$ has its own gauge freedom \cite{Zwanziger1968,Cameron2013,FernandezGuasti2019} to fix the longitudinal part by a choice of gauge. When the radiation gauge ($\nabla\cdot\Crealrt=0$) is chosen, $\Crealrt$ becomes a transverse field, containing the same kind of degrees of freedom as the radiation electromagnetic fields. The electric potential has also been used in the presence of sources  \cite{Cameron2013,Nienhuis2016,FernandezGuasti2019}. The choice of the radiation gauge is also adequate in this case, since it can be shown that the longitudinal degrees of freedom of the field can always be adscribed to the sources instead \cite[I.B.5]{Cohen1997} \cite[Chap.~XXI, \S~22]{Messiah1958}.

In the particular context of integral expressions for the average electromagnetic helicity, the introduction of potentials allows $\rr$-local integrands. This has been shown in the main text in the derivations leading to the $\rr$-local \Eq{eq:helgv}. We will now bring \Eq{eq:AVGtall}, which does not involve the potentials, to the $(\rr,t)$ domain and see how precisely the non-$\rr$-locality arises. We start hence from \Eq{eq:AVGtall}, which only contains the $\Gpplusminus$ fields, and apply the non-result-altering substitutions $\Gpplusminus\rightarrow\Gpplusminus\exp(-ic_0k)$: 
\begin{equation}
	\label{eq:Helapp}
	\begin{split}
		&\AVLambda=\\
		& \intdpconf \left[\Gpplus\exp(-ic_0k)\right]^\dagger \left[\Gpplus\exp(-ic_0k)\right] - \left[\Gpminus\exp(-ic_0k)\right]^\dagger \left[\Gpminus\exp(-ic_0k)\right].
	\end{split}
\end{equation}
We will first focus on the first term of the integrand, which we consider as the $\pp$-point-wise inner product of two functions: $\Gpplus\exp(-ic_0k)$ and $\Gpplus\exp(-ic_0k)/|\pp|$. In order to apply Parseval's theorem and bring \Eq{eq:Helapp} to the $(\rr,t)$ domain, the inverse 3D Fourier transforms of the two functions are needed. We know from the definitions in \Eq{eq:gp} that the inverse 3D Fourier transform of $\Gpplus\exp(-ic_0k)$ is $\Grtplus$. The inverse transform of the product  $\Gpplus\exp(-ic_0k)\times\frac{1}{|\pp|}$ can be obtained by using the convolution theorem and the inverse Fourier transform, denoted by $\FourierinvthreeD{\cdot}$, of each of the two factors (see Eqs~(B.3,B.4) and Tab. II in \cite[I.B.2]{Cohen1997}):
\begin{equation}
	\label{eq:nonloc}
	\FourierinvthreeD{\Gpplus\exp(-ic_0k)\times\frac{1}{|\pp|}}=\frac{1}{2\pi^2}\intdrbar \Grbartplus\times \frac{1}{|\rr-\rrbar|^2}.
\end{equation}
Using \Eq{eq:nonloc} and its obvious counterpart for the second term in the integrand of \Eq{eq:Helapp}, we can use Parseval's theorem to write:
\begin{equation}
	\label{eq:nonlocfinal}
	\AVLambda=\frac{1}{2\pi^2c_0}\intdr\intdrbar \frac{\Grtplus^\dagger\Grbartplus-\Grtminus^\dagger\Grbartminus}{|\rr-\rrbar|^2}.
\end{equation}
As explained in the main text, the typically undesired non-locality of the integrand in \Eq{eq:nonlocfinal} is avoided by the introduction of the helicity potentials, since they absorb the $1/|\pp|$ term\footnote{We note that previously existing non-local expressions for average electromagnetic helicity, like \cite[Eq.~(65)]{Birula2006} and \cite[Eq.~(36)]{Bernabeu2019}, can be shown to be equivalent to \Eq{eq:nonlocfinal}.}. The same kind of arguments show why the introduction of the magnetic and electric potentials, $\Arealrt$ and $\Crealrt$, results in a $\rr$-local integrand in the typical definition of average electromagnetic helicity. Finally, we note that mixed formulations exist where only one of the two potentials is used, which still feature non-$\rr$-local integrands \cite[Eq.~(2.6)]{Deser1976}. 

\section{Equivalence between complex and real versions\label{app:complexreal}}
In this Appendix we show that \Eq{eq:nyaca} of the main text,
\begin{equation}
	\label{eq:nyacaapp}
		\AVLambda=\intdr \frac{1}{Z_0}\Brt^\dagger\Art-Z_0\Drt^\dagger\Crt,
\end{equation}
featuring complex-valued fields is equivalent to the well-known integral for the average helicity featuring real-valued fields, as found e.g. in \cite[Eq.~6]{Crimin2019}:
\begin{equation}
	\label{eq:finalfinal}
	\frac{1}{2}\intdr \frac{1}{Z_0}\Brealrt\cdot\Arealrt-Z_0\Drealrt\cdot\Crealrt.
\end{equation}
We will use properties of complex-valued vector fields whose Fourier transforms contain only positive frequencies, as defined in \Eq{eq:tomega} for $\mathbf{X}$ standing for $\mathbf{A}$, $\mathbf{B}$, $\mathbf{C}$, $\mathbf{D}$, and $\mathbf{E}$: 
\begin{equation}
	\Xrt=\intdomega \Xrw\exp(-i\omega t).
\end{equation}
The real and imaginary parts of $\Xrt=\Xrtreal+i\Xrtimag$ are related by the Hilbert transform, and then their Fourier transforms, denoted by $\Fourier{\cdot}$, meet \cite[p. 49]{Duoandikoetxea2001}:
\begin{equation}
	\label{eq:realim}
	\Fourier{\Xrtimag}=\left(-i\sign\omega\right)\Fourier{\Xrtreal}.
\end{equation}
We now proceed by writing \Eq{eq:nyacaapp} using the real and imaginary parts of each field. Since the end result of the integral must be a real number, we can already discard the imaginary part of the integrand: 
\begin{equation}
	\label{eq:twoterms}
	\begin{split}
		\AVLambda=\intdr&\frac{1}{Z_0}{\Brtreal}^\dagger\Artreal+\frac{1}{Z_0}{\Brtimag}^\dagger\Artimag\\
		&-Z_0{\Drtreal}^\dagger\Crtreal-Z_0{\Drtimag}^\dagger\Crtimag=\\
		\intdr& \left[\frac{1}{Z_0}{\Brtreal}^\dagger\Artreal-Z_0{\Drtreal}^\dagger\Crtreal\right]+\\
		& \left[\frac{1}{Z_0}{\Brtimag}^\dagger\Artimag-Z_0{\Drtimag}^\dagger\Crtimag\right].
	\end{split}
\end{equation}
We will now show that the two expressions in square brackets produce the same contribution. To such end, let us focus on one of their terms and use the time-harmonic decomposition\footnote{The one sided integral in \Eq{eq:real} can be written as a two sided integral over the frequency axis and the familiar result ${\Xrw}^*=\mathbf{X}^{-\omega}(\rr)$ is recovered.}
\begin{equation}
	\begin{split}
		&\intdr {\Brtimag}^\dagger\Artimag=\\
		&\intdr \left[\intdomegatwo \Fourier{\Brtimag}\exp(-i\omega t)\right]^{\dagger}\left[\intdomegatwo \Fourier{\Artimag}\exp(-i\omega t)\right].
	\end{split}
\end{equation}
The same considerations that take the last line of \Eq{eq:jarl} to the first line of \Eq{eq:lots} can be used to write:
\begin{equation}
	\label{eq:needthis}
	(2\pi)\intdr {\Brtimag}^\dagger\Artimag=\intdr \intdomegatwonopi {\Fourier{\Brtimag}}^\dagger\Fourier{\Artimag}.
\end{equation}
We now use \Eq{eq:realim} on the last expression in \Eq{eq:needthis}
\begin{equation}
	\label{eq:ais}
	\begin{split}
	&(2\pi)\intdr {\Brtimag}^\dagger\Artimag=\\
		&\intdr \intdomegatwonopi \left[\left(-i\sign\omega\right){\Fourier{\Brtreal}}\right]^\dagger\left[\left(-i\sign\omega\right)\Fourier{\Artreal}\right]=\\
		&\intdr \intdomegatwonopi {\Fourier{\Brtreal}}^\dagger\Fourier{\Artreal}=(2\pi)\intdr {\Brtreal}^\dagger\Artreal,\\
	\end{split}
\end{equation}
where the last equality follows by comparison with \Eq{eq:needthis}. Equation~(\ref{eq:ais}) shows that the two expressions inside the square brackets in \Eq{eq:twoterms} produce the same contribution, since the steps leading to \Eq{eq:ais} can be applied to any of the product terms. We can hence write:
\begin{equation}
	\label{eq:ff}
		\AVLambda=2\intdr\frac{1}{Z_0}{\Brtreal}^\dagger\Artreal- Z_0\Drtreal^\dagger\Crtreal.
\end{equation}
Equivalence with \Eq{eq:finalfinal} is shown after considering that the definition of the typical real fields $\Xtildert$ in \Eq{eq:real} implies that $\Xtildert=2\Xrtreal$. Equation~(\ref{eq:finalfinal}) is finally reached by substituting all the $\Xrtreal$ fields with $\Xtildert/2$ in \Eq{eq:ff}.

\section{Expressions for computing Circular Dichroism\label{app:cd}}
Let us assume that a chiral molecule is located at point $\rr$, and embedded in a possibly frequency-dispersive, homogeneous, isotropic, achiral, and non-magnetic medium with permittivity $\epsilon_m^\omega$, permeability $\mu_m^\omega=\mu_0$, impedance $Z^\omega=\sqrt{\mu_0/\epsilon^\omega_m}$, speed of light $c_\omega=1/\sqrt{\mu_0\epsilon^\omega_m}$. The wavenumber in such medium is $k_\omega=\omega/c_\omega$. In this Appendix, these frequency-dependent quantities are assumed to substitute their constant vacuum counterparts in all the equations in the main text. 

The most common time-harmonic light-molecule interaction model is the 6$\times$6 dipolarizability tensor, which relates the external electric and magnetic fields at the location of the molecule with the electric [$\textbf{p}^\omega(\rr)$] and magnetic [$\textbf{m}^\omega(\rr)$] dipoles induced by the fields in the molecule:
\begin{equation}
	\label{eq:dipoles}
\begin{bmatrix} \textbf{p}^\omega(\rr) \\ \textbf{m}^\omega(\rr) \end{bmatrix} =\alphatensor\begin{bmatrix}\textbf{E}^\omega(\rr) \\ \textbf{H}^\omega(\rr) \end{bmatrix}.
\end{equation}
One of the most relevant techniques for chiral molecule sensing is Circular Dichroism (CD), which measures their differential absorption upon subsequent illumination with the two helicities. Assuming that the field scattered by the molecule, i.e., the field radiated by the induced dipoles in \Eq{eq:dipoles}, is negligible with respect to the incident field, it is possible to write the rotationally averaged molecular differential absorption as \cite{Graf2019}.
\begin{equation}
	\label{eq:Aalpha}
	\text{CD}(\rr)=\intdomeganopi \alphaemp\frac{\omega c_\omega}{2}\left[|\Grwplus|^2-|\Grwminus|^2\right]=\intdomeganopi \alphaemp 2c_\omega^2 C^\omega(\rr),
\end{equation}
where $\alphaemp$ is the real part of the rotational average of $\matr{\alpha}^\omega_\text{me}$, $C^\omega(\rr)$ is the optical chirality density introduced by Tang and Cohen \cite{Tang2010}, and the second equality follows from \cite[Eq.~(5)]{FerCor2016} and \Eq{eq:gp}.

Besides the dipolarizability model in \Eq{eq:dipoles}, other light-molecule interaction descriptions are possible. For example, the T-matrix of the molecule may be used. The T-matrix is a common object in physics and engineering, which is intrinsically able to include all the multipolar orders of the light-matter interaction, and allows to efficiently compute the coupled electromagnetic response of different objects in a systematic and rigorous way \cite{Mishchenko2017}. The conversion between the dipolarizability tensor and the T-matrix of the molecule up to the dipolar order is \cite[Eq.~(A15)]{FerCor2018}:
\begin{equation}
	\alphatensor=\frac{-i6\pi}{c_\omega Z_\omega k_\omega^3}\begin{bmatrix}\matr{T}^\omega_{NN}&iZ\matr{T}^\omega_{NM}\\-ic_\omega\matr{T}^\omega_{MN}&c_\omega Z\matr{T}^\omega_{MM}\end{bmatrix},
\end{equation}
where $N(M)$ refers to the electric(magnetic) character. We may now use this conversion to substitute $\alphaemp$ in \Eq{eq:Aalpha}:
\begin{equation}
	\label{eq:CDFV}
	\begin{split}
		\text{CD}(\rr)&=\intdomeganopi \Tnmp(3\pi c_\omega ^3)\frac{|\Grwplus|^2-|\Grwminus|^2}{\omega^2}\\
		&=\intdomeganopi\Tnmp(3\pi c_\omega^3)\left[|\Vrwplus|^2-|\Vrwminus|^2\right].
	\end{split}
\end{equation}
In experimental measurements, a solution of chiral molecules is confined in its recipient, which defines a volume $\mathbb{D}$. Assuming uniform concentration of molecules over $\mathbb{D}$, the total CD signal can be computed by the volume integral of any of the expressions in \Eq{eq:Aalpha} or \Eq{eq:CDFV} over $\mathbb{D}$. For example:
\begin{equation}
	\label{eq:CD}
	\begin{split}
		\text{CD}&=\rho\intdrD\intdomeganopi \Tnmp(3\pi c_\omega^3)\frac{|\Grwplus|^2-|\Grwminus|^2}{\omega^2}\\&=\rho\intdomeganopi \frac{\Tnmp(3\pi c_\omega^3)}{\omega}\intdrD \frac{|\Grwplus|^2-|\Grwminus|^2}{\omega},
	\end{split}
\end{equation}
where $\rho$ is a constant that depends on the molecular concentration. In the last expression, we recognize one of the integrands from the average helicity in \Eq{eq:lots}. 
\clearpage{}
\end{document}